\documentclass[twocolumn,showpacs,preprintnumbers,amsmath,amssymb]{revtex4}

\usepackage{graphicx}
\usepackage{dcolumn}
\usepackage{bm}

\begin{document}

\title{Power Dependence of the Photocurrent Lineshape in a Semiconductor Quantum Dot}

\author{A. Russell}
\author{Vladimir I. Fal'ko}
\affiliation{Department of Physics, University of Lancaster,
Lancaster, LA1 4YB, UK}

\date{\today}

\begin{abstract}
We propose a kinetic theory to describe the power dependence,
$I_{PC}(P)$, of the photocurrent (PC) lineshape in optically pumped
quantum dots at low temperatures, in both zero and finite magnetic
fields. We show that there is a crossover power $P_c$, determined by
the electron and hole tunneling rates, at which the photocurrent
spectra become strongly influenced by the dot kinetics, and no
longer reflect the exciton lifetime in the dot. For $P>P_c$, we show
that the photocurrent saturates due to the slow hole escape rate (in
e.g., InGaAs/GaAs dots), whereas the line-width increases with
power: $\Gamma \propto \sqrt{P}$. We also analyze to what measure
the spin-doublet lineshape of the photocurrent studied in a high
magnetic field reflects the degree of circular polarization of the
incident light.
\end{abstract}


\maketitle

In the recent years, advances in experimental techniques have
allowed for optical studies on single self-assembled quantum dots,
e.g., by using spatially resolved photocurrent (PC) spectroscopy
\cite{chang, findeis,beham,oulton,simma,hoglund}. Typically in these
experiments, quantum dots (QDs) were embedded in biassed PIN
junctions, permitting to control the tunneling rates of optically
generated electrons and holes from the dot and, therefore, the
photocurrent formed by electrons and holes escaping in opposite
directions. In principle, measurements of photocurrent can be used
as a detection method to study spectral properties of a dot
\cite{fry,oulton2,oulton3,pettersson,stufler}. In this Letter, we
analyse the kinetics of an optically pumped quantum dot, such as a
InGaAs/GaAs QD, and study the power dependence of the photocurrent.
We show that although at low powers, the PC line-width manifests the
exciton life-time in the dot, at high powers it becomes strongly
influenced by dot kinetics and no longer reflects the properties of
the dot. The crossover power between these two regimes is found to
depend on both the electron and hole tunneling rates, as well as the
efficiency with which photons are absorbed by the QD. The analysis
is also extended onto the case of high magnetic fields and takes
into account the degree of circular polarization of the incident
light.

\begin{figure}
\centering
\includegraphics[width=8cm]{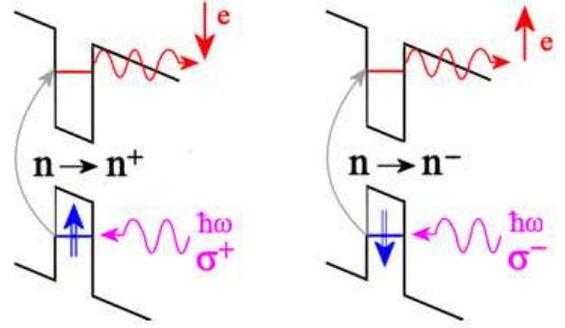}
\caption{On the left, a schematic of the dot showing the processes
involved in the rate $w_{+}$. The dot is excited by a $\sigma^+$
photon, generating an exciton consisting of a spin
$J_z=+\tfrac{3}{2}$ hole and a spin $S_z = -\tfrac{1}{2}$ electron,
which quickly tunnels to the contacts resulting in the dot being
positively charged. On the right, the same but for the rate $w_{-}$
corresponding to excitation by a $\sigma^-$ photon.} \label{fig1}
\end{figure}

To model the power dependence of the photocurrent, $I_{PC}$,
generated when a quantum dot is optically pumped with monochromatic
light of frequency $\omega$ and power $P$, we consider the processes
shown in Fig.~\ref{fig1}. Here, a pair of an electron and heavy hole
in the lowest confined states in the dot (exciton) is created by
either a $\sigma^{+}$ or a $\sigma^{-}$ photon, followed by their
escape to an electrode. In large electric fields $F$, the dominant
escape mechanism from the dot for the electron and hole is by
tunneling, with the power-independent rates $\gamma_{e,h}/\hbar
\propto \exp(-4 \sqrt {2m_{e,h}\Delta V^{3}_{e,h}}  / 3 e \hbar F)$
\cite{landau,gamma,larkin}. Due to a smaller electron mass, it is
natural to assume that the hole tunnels slower than the electron
($\gamma_e \gg \gamma_h$). The escape of a photo-excited electron
into the bulk semiconductor (e.g., GaAs) permits slightly
off-resonant excitation of the dot \cite{one}, where the energy of
the absorbed photon, $\hbar \omega$, slightly differs from the
exciton energy, $\epsilon_0$. Note that in such a case, energy is
conserved by the electron tunneling into the bulk semiconductor
continuum of energy states, with the exciton appearing on the dot
only virtually. As a result, the QD tends to be positively charged,
blocking the absorption of another photon until the hole escapes and
thus limiting the size of the photocurrent (which is generated by
the repetition of such processes). Blocking of the next inter-band
transition on the dot happens due to (a) the Fermi blocking for the
transition in the same circular polarization and (b) the strong
shift of the inter-band transition energy for a transition in the
opposite polarization exciting the dot into a charged exciton state
(as compared to the neutral one \cite{kuo,cook,two}).

Below we develop a theory for the PC spectroscopy of dots in both
zero and finite magnetic field. The latter gives rise to an exciton
Zeeman splitting $\epsilon_{Z}^{x}$, making the dot kinetics
sensitive to the polarization degree of the optical pump, $\sigma$,
where $\sigma=\pm 1$ corresponds to fully polarized $\sigma^{\pm}$
light. The projection of the photon angular momentum onto the
$z$-axis, defined as the growth direction of the dot, is conserved
in the system. Due to this, $\sigma^{\pm}$ photons can only excite
spin $S_{z}=\mp \tfrac{1}{2}$ electrons and spin $J_{z}= \pm
\tfrac{3}{2}$ holes, forming excitons of energy $\epsilon_{0}\pm
\tfrac{1}{2}\epsilon_{Z}^{x}$, Fig.~\ref{fig1}. The excitation rate
of the empty dot into a positively charged state with a
$J_{z}=\pm\tfrac{3}{2}$ hole takes a resonant form,
\begin{equation}
w_{\pm}=\frac{1 \pm \sigma}{2}\frac{\gamma_e \alpha P}{\left(\delta
 \pm \tfrac{1}{2}\epsilon_{Z}^{x}\right)^2 + \tfrac{1}{4}\gamma_e^2},\qquad
\delta = \hbar\omega - \epsilon_{0}. \label{rate}
\end{equation}
The constant $\alpha$ characterizes oscillator strength of the
optical transition and is defined by $|A_{eh}|^{2}=\alpha P$, where
$A_{eh}$ is the interband transition amplitude in the presence of
the electromagnetic field of light.

\begin{figure}
\centering
\includegraphics[width=8cm]{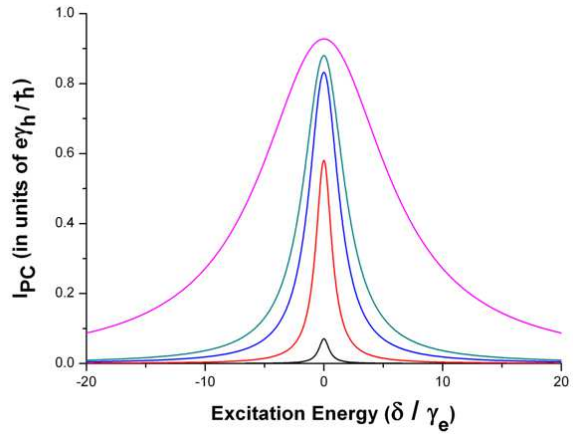}
\caption{The photocurrent, $I_{PC}$, in the case of zero external
magnetic field as a function of excitation energy for different
powers $P/P_c= 0.1,1,5,20,200$ (from the bottom to the top).}
\label{fig2}
\end{figure}

Below, we use $n$ to represent the probability that the dot is
empty, and $n^{\pm}$ for it to be occupied by a
$J_{z}=\pm\tfrac{3}{2}$ hole \cite{two}. The balance equations for
the positively charged states of the dot have the following form,
\begin{equation}
\dot{n}^{\pm} = w_{\pm}n - \frac{\gamma_h}{\hbar}n^{\pm},
\label{balance}
\end{equation}
where the first term describes the excitation of the dot into a
positively charged state, and the last term represents the hole
escape. We solve Eq.~(\ref{balance}) for the steady state by
supplementing them with the normalization condition $n +
n^{+}+n^{-}=1$, and find that
\begin{equation}
n=\frac{\gamma_h}{\gamma_h + \hbar\left( w_{-} + w_{+}\right)}.
\label{empty}
\end{equation}
This determines the photocurrent,
\begin{equation}
I_{PC}=e(w_{+} + w_{-})n=\frac{e\left(w_{-} +
w_{+}\right)\gamma_h}{\gamma_h + \hbar\left(w_- + w_+ \right)}.
\label{IPCfull}
\end{equation}
Using Eq.~(\ref{rate}) we can now describe the photocurrent as a
function of the electron and hole tunneling rates
($\gamma_e,\gamma_h$), and the power, frequency and polarization
degree of light ($P,\omega,\sigma$). In zero magnetic field, the
photocurrent, given by Eq.~(\ref{IPCfull}), loses its $\sigma$
dependence and only has one resonance at $\delta=0$,
\begin{equation}
I_{PC}=\frac{e\gamma_e\alpha P}{\delta^2 + \Gamma^2},
\label{IPCsimple}
\end{equation}
where, typically for two-level systems \cite{textbook},
\begin{equation}
\Gamma=\frac{1}{2}\gamma_e
\sqrt{1+\frac{P}{P_c}},\;\;\mathrm{where}\;\;
P_c=\frac{\gamma_e\gamma_h}{4\alpha \hbar}. \label{gamma}
\end{equation}
The lineshape of the dot as manifested in the photocurrent,
Eq.~(\ref{IPCsimple}), shown in Fig.~\ref{fig2} for various
different powers. It is characterized by the line-width $\Gamma$ and
integral strength $A=\int{I_{PC}d\omega}$,
\begin{equation}
A=\frac{ 2e\pi \alpha P}{\hbar \sqrt{1+P/P_c} }. \label{charact}
\end{equation}
(The latter should not be confused with the integral photocurrent
generated by white light \cite{integral}.)

The properties of the QD spectrum measured using photocurrent can be
described by considering the two power regimes $P\gtreqless P_c$. In
low powers, $P<P_c$, the peak height, $I_{PC}(\delta=0)$, increases
proportionally to power, and the line-width, $\Gamma \propto
\tfrac{1}{2}\gamma_e$, reflects the lifetime of the exciton in the
dot. When $P > P_c$, the line-width becomes strongly
power-dependent: $\Gamma \sim \tfrac{1}{2}\gamma_e\sqrt{P/P_c}$.
This result is similar to the phenomenological anzats made in
Ref.~\cite{stufler} and agrees with the recent observations in Ref.
\cite{oulton3}. Simultaneously, the integral strength increases at
high powers, $A \approx \sqrt{P}$, whereas the peak height saturates
at $e\gamma_h/\hbar$, indicating that the slow hole tunneling rate
limits the magnitude of the photocurrent (as observed in
Refs.~\cite{beham,stufler,oulton3}).

In the presence of a magnetic field, photo-excitation of the dot
with fully polarized light ($\sigma=\pm 1$) results in the same
photocurrent properties as described above, since only an exciton of
one polarization can be excited into the dot. For $\sigma \neq \pm
1$, the photocurrent takes the bi-modal form shown in
Fig.~\ref{fig2} for $\sigma = 0.6$. In Fig.~\ref{fig2}(a), the
Zeeman splitting is small, and even at low powers the two peaks
overlap. In the high power limit, $P \gg P_c$, the photocurrent
behaves in a similar fashion as in the zero magnetic field case.

\begin{figure}
\centering
\includegraphics[width=8cm]{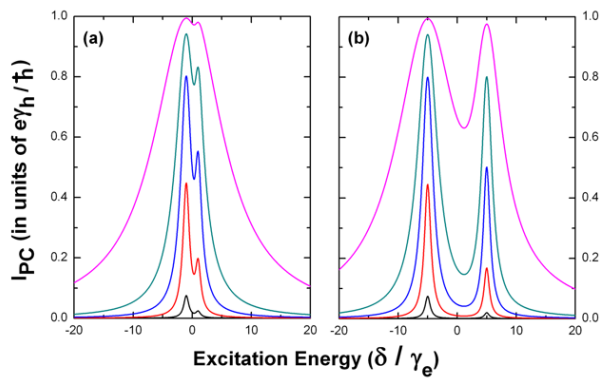}
\caption{(a). Photocurrent spectra given by Eq.~(\ref{IPCfull}) for
$\sigma=0.6$, exciton Zeeman energy $\epsilon_{Z}^{x}=2\gamma_e$ and
powers $P/P_c= 0.1,1,5,20,200$ (from the bottom to the top); (b).
The same, but for $\epsilon_{Z}^{x}=10\gamma_e$.}\label{fig3}
\end{figure}

If $\epsilon_{Z}^{x}$ is large as compared to a typical line-width,
the two photocurrent peaks become distinct even at high powers,
Fig.~\ref{fig2}(b), and both display similar properties as the
single peak described in Eq.~(\ref{IPCsimple}). The value of
$\sigma$ determines how often one electron-hole spin configuration
is excited into the dot as compared to the other. Thus one may
expect that the polarization degree of light is reflected in the
magnitude of the photocurrent at the energies $\delta \pm
\tfrac{1}{2} \epsilon_{Z}^{x}$, as is the case for low powers seen
in Fig.~\ref{fig2}(b). However, for high powers $P\gg P_c$, the
peaks both saturate at $e\gamma_h/\hbar$ as described previously,
and therefore their heights no longer reflect the value of $\sigma$.
Instead, it is now manifested by their respective line-widths which
reflect the size of the area under the corresponding line.

To conclude, we have shown that there is a crossover power $P_c$
beyond which the line-width of the photocurrent no longer manifests
the lifetime of electron states on the dot. At low powers, the dot
(such as an InGaAs dot in a GaAs matrix) is nearly always empty and
the photocurrent is proportional to the excitation rate, with
broadening then reflecting the lifetime of the exciton. At high
powers, the dot is more often occupied by a hole (therefore it is
positively charged) and the height of the photocurrent peak height
saturates at $e\gamma_h/\hbar$, determined by the slower hole escape
rate. The crossover power, $P_c=\gamma_e\gamma_h/4\hbar\alpha$,
depends on both the electron and hole escape rates, as well as the
efficiency with which the dot absorbs photons incident on the dot.
The proposed theory of PC lineshape at low temperatures explains the
recent experimental observations of PC measurements on optically
pumped quantum dots \cite{beham,oulton3,stufler}.

We thank A. I. Tartakovskii and M. S. Skolnick for stimulating
discussions. This work has been funded by two ESF-EPSRC FoNE
projects: SpiCo and SPINCURRENT.


\begin{thebibliography}{}
\bibitem{gamma}We ignore the small affects on $\gamma_{e,h}$ due to
the electron and hole Zeeman energies, so that $\Delta V_{e,h}$ is
independent of the electron/hole spin.
\bibitem{one}In Ref. \cite{beham}, a kinetic model has been used
which considered resonant excitation of the dot, where the dot can
be occupied by both a photo-excited electron and a hole. Such a
model is formally applicable only when inelastic (e.g.,
electron-phonon) interactionsmakes it possible for the optical
absorption of a non-resonant photon, which may be the case at high
temperatures. Also, the analysis in Ref. \cite{beham} does not give
the power dependence of the lineshape of the photocurrent. In
contrast, the analysis presented in this paper, which focusses on
the low temperature regime, is applicable to the description of the
PC generated by slightly off-resonant photons. This is because, in
the presented analysis, an exciton occupies the dot only virtually,
with energy conserved by the electron tunneling into the spectral
continuum of the bulk semiconductor. As such our theory not only
describes the saturation of the peak height, but also predicts the
lineshape of the photocurrent, and the power broadening as
experimentally shown in Refs. \cite{oulton3,stufler}.
\bibitem{two}Double occupation of the dot by
two holes is prohibitted in our model by a large Coulomb energy,
$E_c \sim \tfrac{e^2}{\epsilon a} \gg T$, where $\epsilon$ is the
dielectric constant and $a$ is the size of the dot.
\bibitem{integral}The integral photocurrent is generated by
non-monochromatic light which one would attribute to a QD assemble,
pumping the quantum dot with photons of various frequencies. In such
a case, the excitation rate in Eq.~(\ref{rate}) is given by the
integral,
\begin{equation}
w_{\pm}=\frac{1\pm \sigma}{2}\int{\frac{e\gamma_e \alpha
\rho(\omega)}{{\left(\delta \pm
\tfrac{1}{2}\epsilon_{Z}^{x}\right)^2 +
\tfrac{1}{4}\gamma_e^2}}d\omega}.\nonumber
\end{equation}
Here $\rho(\omega)$ is the spectral density of photons arriving with
frequency $\omega$, with the units of rate. Integrating, and
substituting into Eq.~(\ref{IPCfull}) we arrive at
\begin{equation}
I_{int}=\frac{2e\alpha P}{1+\tfrac{1}{2}\gamma_e P/P_c}, \nonumber
\end{equation}
in agreement with Refs. \cite{beham,oulton3,stufler}.
\end{thebibliography}
\end{document}